\documentclass[pra, superscriptaddress,twocolumn,showpacs,nofootinbibfloatfix,amsmath,amsfonts,amssymb]{revtex4-1}%
\usepackage{amsmath,amsfonts,amssymb,color}
\usepackage{amsthm}
\usepackage{leftidx}
\usepackage{graphicx}
\usepackage{xcolor}
\usepackage{dcolumn}
\usepackage{bm}
\usepackage{epstopdf}
\usepackage{epsfig}

\begin{document}
	\title{Combating quasiparticle poisoning with multiple Majorana fermions in a periodically-driven quantum wire}
	\author{Raditya Weda Bomantara}
	\email{Raditya.Bomantara@sydney,edu,au}
	\affiliation{%
		Department of Physics, National University of Singapore, Singapore 117543
	}
	\affiliation{%
	Centre for Engineered Quantum Systems, School of Physics, University of Sydney, Sydney, New South Wales 2006, Australia
	}
	\author{Jiangbin Gong}%
	\email{phygj@nus.edu.sg}
	\affiliation{%
		Department of Physics, National University of Singapore, Singapore 117543
	}
	\date{\today}
	
	
	\vspace{2cm}
	
\begin{abstract}
	Quasiparticle poisoning has remained one of the main challenges in the implementation of Majorana-based quantum computing. It inevitably occurs when the system hosting Majorana qubits is not completely isolated from its surrounding, thus considerably limiting its computational time. We propose the use of periodic driving to generate multiple MFs at each end of a single quantum wire, which naturally provides the necessary resources to implement active quantum error corrections with minimal space overhead. In particular, we present a stabilizer code protocol that can specifically detect and correct any single quasiparticle poisoning event. Such a protocol is implementable via existing proximitized semiconducting nanowire proposals, where all of its stabilizer operators can be measured from an appropriate Majorana parity dependent four-terminal conductance.
	
\end{abstract}

\maketitle

\section{Introduction}
\label{intro}

The quest for achieving fault-tolerant quantum computing remains an ongoing research direction since over the last two decades. Indeed, one main challenge in the realization of large-scale quantum computers is their sensitivity to noise and system imperfections. It is estimated that even with the implementation of quantum error correction, an error threshold of $\sim 10^{-4}$ is necessary to ensure properly functioning quantum information processing devices \cite{knill,thre}.

The advent of topological phases of matter has led to the idea of utilizing topological degree of freedom to establish fault-tolerance through the so-called topological quantum computing (TQC) \cite{tqc,tqc2}, which was originally developed in Ref.~\cite{Kitqc}. Within the framework of TQC, logical qubits are encoded in the Hilbert space of anyons, i.e., exotic quasiparticles living in two-dimensions (2D) that usually arise in defects of certain topological materials, and quantum gate operations are accomplished by adiabatically moving one anyon around another (a process termed braiding). The robustness of anyon-based qubits originates from the nonlocality that arises when a pair of anyons that are very far apart are used to encode such qubits, whereas the robustness of braiding owes to the fact that its effect on the logical Hilbert space is insensitive to the adiabatic path that realizes such a process.

In recent years, the area of TQC has been actively pursued, whose main focus revolves around the use of a specific kind of anyons called Majorana fermions (MFs). Among other types of anyons, MFs are currently considered to be the most promising candidates for the near future experimental implementation of TQC due to the abundant realistic proposals for creating MFs in cold-atoms \cite{opt} and proximitized semiconductors \cite{semi1,semi2} or topological insulator edges \cite{MTI}, as well as the various braiding proposals in these platforms \cite{wire,Tjun1,Tjun2,braid4,mea3,braid5,braid6,braid7}. Moreover, experimental signatures of MFs have been confirmed recently \cite{exp1,exp2,exp3}, which opens up possibilities to experimentally realize Majorana-based TQC in the near future.

While Majorana-based qubits are expected to enjoy topological protection as pointed out above, their lifetime is limited in present experiments. In particular, since their realizations involve a contact between different materials, a flow of quasiparticles between them is inevitable and may lead to errors termed `quasiparticle poisoning' (QP) \cite{QP,QP2,QP3,QP4,QP5}. In the presence of QP, it is estimated that the lifetime of Majorana-based qubits ranges between $10$ ns and $0.1$ ms \cite{QP}, which may pose problems in the implementation of adiabatic-based braiding proposals of Refs.~\cite{wire,Tjun1,Tjun2,braid4,braid5,braid6,braid7}.

The presence of charging energy, accomplished by connecting a topological superconductor with an external capacitor, is known to suppress QP rates, { and consequently enhance the qubits' lifetime}. Measurement-based braiding proposals of Refs.~\cite{mea,mea1,mea2,mea3,mea4,mea5,mea6,QP2,surcode,RG}, { whose implementation necessarily involves a finite charging energy in the system}, are thus expected to be more resistant to QP errors and may be easier to implement experimentally. Nevertheless, at longer computational times, QP errors may become significant, and complementing these measurement-based quantum information processing devices with active quantum error corrections is necessary.

In this paper, we utilize a feature of periodically-driven (Floquet) topological phases to naturally implement a quantum error correction code. In particular, unlike static topological phases which are generically only able to host a fixed number of edge states at their boundaries, certain Floquet topological phases can be designed to host multiple edge states, whose number can be increased without bound by tuning some system parameters \cite{LW,LW2,LW3,RG2}. In the context of topological superconductors, the use of periodic driving then allows a single one-dimensional (1D) quantum wire to host as many MFs as possible. This possibility is first demonstrated in Ref.~\cite{kk3}, or alternatively one may map the results of Refs.~\cite{LW,LW2,LW3,RG2} to topological superconductors and generate many MFs in a more controllable manner.

Here, we propose an even simpler recipe to generate many MFs in a Floquet quantum wire that does not involve changing the superconducting pairing phase \cite{kk3} or suppressing all but one parameters to zero \cite{LW,LW2,LW3,RG2,LW4}. Moreover, parameter regime at which any particular (even) number of MFs exist in the system can be analytically identified, which thus in principle allows the implementation of various quantum error correction codes. In particular, we will present an example of such codes with the capability of detecting and correcting any single QP error, which can be implemented via conductance measurements and is thus compatible with existing measurement-based Majorana qubit architectures. {  It is noted that the typical measurement-times associated with the aforementioned conductance readout are of the order of tens of nanoseconds \cite{mea6}, which are well below the Majorana qubits lifetime above, thus justifying the use of such conductance measurement-based error correction codes.}

This paper is structured as follows. In Sec.~\ref{sec2}, we introduce our proposed Floquet quantum wire model which allows multiple MFs to be realized at its ends by tuning some system parameters. In Sec.~\ref{det}, we present a possible implementation of the Majorana Steane $[[7,1,3]]$ code with our proposed model by appropriately tuning the system parameters within a certain regime. In Sec.~\ref{condmea}, we elucidate a possible physical realization of stabilizer measurements via parity-dependent conductance. In Sec.~\ref{scale}, we discuss a means to integrate our model into scalable designs proposed in Ref.~\cite{QP2}, thus opening up a possibility to encode and manipulate many logical qubits in a topologically protected manner, each of which is equipped with an additional error correction code that further protects such qubits against QP errors. Finally, we summarize our paper in Ref.~\ref{conc} and discuss some potential future directions.

\section{Generating many Majorana Fermions in a 1D quantum wire}
\label{sec2}

In this section, we present a simple recipe for generating many MFs in a periodically driven 1D quantum wire. To this end, we consider the Hamiltonian
\begin{equation}
H(t) = \sum_j \left(\frac{\mu(t)}{2} c_j^\dagger c_j -J(t) c_{j+1}^\dagger c_j +\Delta(t) c_{j+1}^\dagger c_j^\dagger\right) +h.c. \;, \label{mod}
\end{equation}
where $\mu(t)$, $J(t)$, and $\Delta(t)$ are the time-periodic chemical potential, hopping, and pairing amplitudes respectively (all taken to be reals), $c_j$ ($c_j^\dagger$) is the fermionic annihilation (creation) operator at site $j$. In this paper, we consider a binary time-periodic drive, so that $H(t)=H_i$ with $\mu(t)=\mu_i$, $J(t)=J_i$, and $\Delta(t)=\Delta_i$ for $(n+(i-1)/2)T<t\leq (n+i/2)T$, where $n$ is an integer, $T=\frac{2\pi}{\omega}$ is the period of the drive, and $i=1,2$. That is, the system switches between two Kitaev Hamiltonian \cite{Kit} within one period.


To understand how such a simple system can host many MFs in a controllable manner, we may first take periodic boundary conditions (PBC) and write Eq.~(\ref{mod}) in momentum space as
\begin{equation}
H(t) = \frac{1}{2} \sum_k \psi_k^\dagger h(k,t) \psi_k\;,
\label{mom1}
\end{equation}
where
\begin{equation}
h(k,t) = [\mu(t) - 2J(t) \cos(k)] \sigma_z + 2\Delta(t) \sin(k) \sigma_y \;,\label{mom}
\end{equation}
$\psi_k=\left(c_k,\; c_{-k}^\dagger\right)^T$ is a vector in the Nambu space, $\sigma_i$ are the associated Pauli matrices, and we have set the lattice constant to unity. Since $h(k,t)$ is time-periodic, its topology is encoded by its quasienergy bands $\varepsilon_n (k)\in(-\pi/T,\pi/T]$, defined as the eigenphase of the one-period evolution operator (termed Floquet operator onwards) \cite{Flo1,Flo2}, i.e.,
\begin{equation}
U(k,t_0+T;t_0)|\varepsilon_n\rangle = e^{-\mathrm{i} \varepsilon_n T/\hbar} |\varepsilon_n\rangle \;.
\end{equation}

For a binary drive considered in this paper, such a Floquet operator can be readily obtained as a product of exponentials as (taking $t_0=T/4$ without loss of generality)
\begin{eqnarray}
U(k)&\equiv& U(k,t+T;t) =F(k)G(k) \;, \nonumber \\
F(k) &=&  \exp\left(-\mathrm{i} \frac{h_1 T}{4\hbar}\right) \times \exp\left(-\mathrm{i} \frac{h_2 T}{4\hbar}\right) \;, \nonumber \\
G(k) &=& \exp\left(-\mathrm{i} \frac{h_2 T}{4\hbar}\right) \times \exp\left(-\mathrm{i} \frac{h_1 T}{4\hbar}\right) \;, \label{stf}
\end{eqnarray}
where $h_1(k)$ and $h_2(k)$ are the momentum space Hamiltonian associated with $H_1$ and $H_2$ respectively (see Eqs.~(\ref{mom1}) and (\ref{mom})). In such a symmetric time frame \cite{TSF,TSF3,DG}, it is also easy to verify that our system exhibits a chiral symmetry under the operator $\Gamma= \sigma_x$, which maps $\Gamma F(k) \Gamma^\dagger = G(k)^\dagger$. In the presence of chiral symmetry, any quasienergy eigenstate $|\varepsilon\rangle$ implies the existence of another quasienergy eigenstate $|-\varepsilon\rangle = \Gamma |\varepsilon\rangle$. Under open boundary conditions (OBC), this implies the protection of degenerate edge states at quasienergy $0$ and $\pi/T$, which are respectively referred to as Majorana zero modes (MZMs) and Majorana $\pi$ modes (MPMs).

As discussed in \cite{TSF,TSF3}, in such a chiral-symmetric Floquet system, two winding number invariants $\nu_0$ and $\nu_\pi$ can be defined, which in our case count the number of MZMs and MPMs existing in our system for a given set of parameter values. To this end, we first recast $F(k)$ in a canonical ($\Gamma\rightarrow \sigma_z$) basis as
\begin{equation}
F(k) \hat{=}\left( \begin{array}{cc}
A(k) & B(k) \\ C(k) & D(k)
\end{array} \right)  \;,
\end{equation}
after which we can define these winding numbers as \cite{TSF,TSF3}
\begin{eqnarray}
\nu_0 &=& \frac{1}{2\pi\mathrm{i}}\int dk\;  B^{-1} \frac{dB}{dk} \;, \nonumber \\
\nu_\pi &=&   \frac{1}{2\pi\mathrm{i}}\int dk\;  D^{-1} \frac{dD}{dk} \;.
\end{eqnarray}

\begin{figure}
	\centering
	\includegraphics[scale=0.25]{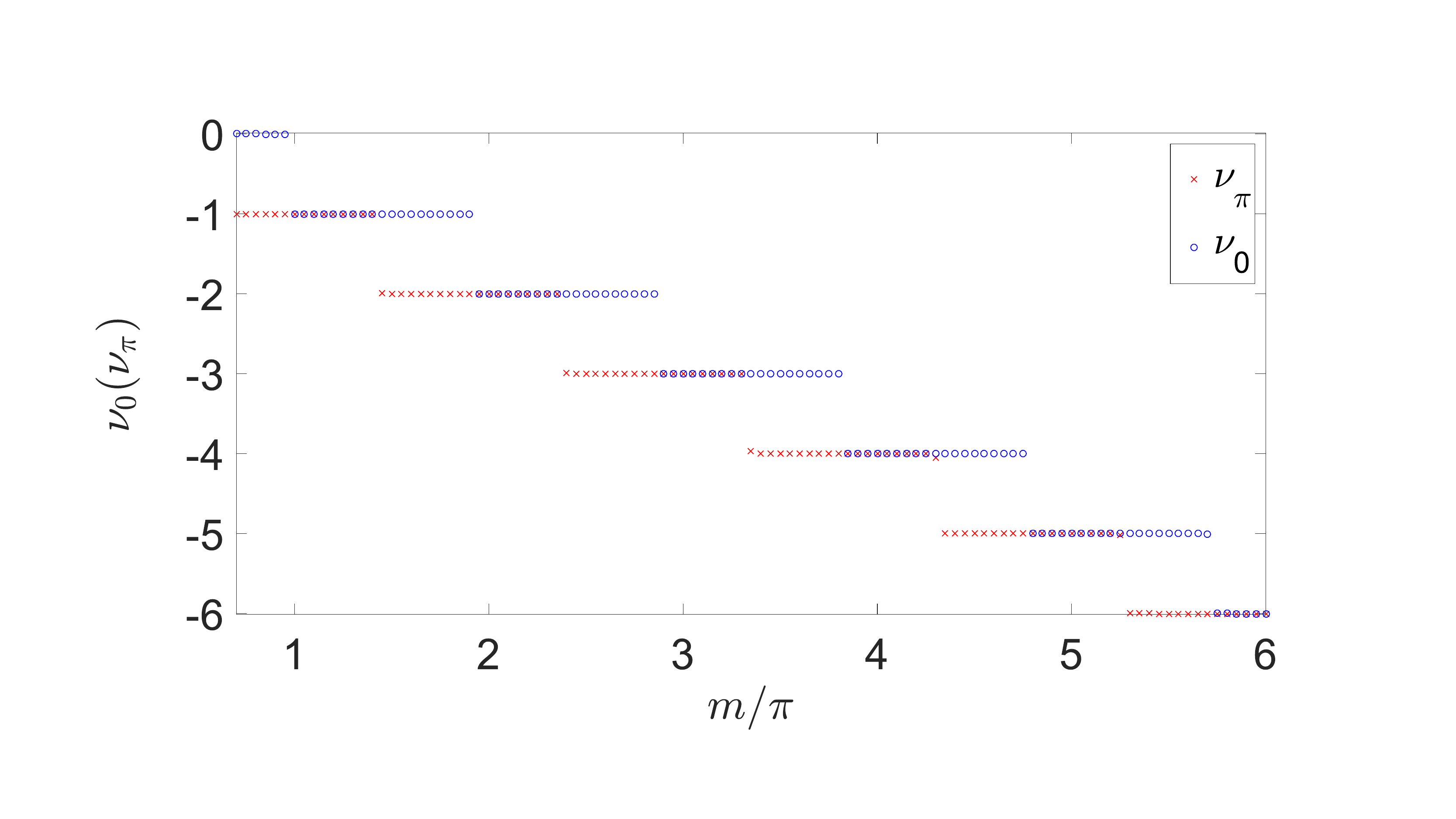}
	\caption{Winding number invariants $\nu_0$ and $\nu_\pi$ as the parameter $m$ (see main text) is increased. Here, we take $\frac{\mu_1 T}{2\hbar}=1$, $\frac{J_1 T}{2\hbar}=0.55$, and $\frac{\Delta_1 T}{2\hbar}=0.6$.}
	\label{wmj}
\end{figure}

Figure~\ref{wmj} shows the winding number invariants $\nu_0$ and $\nu_\pi$ as some parameters are varied according to $\mu_2=m\mu_1$, $J_2=-mJ_1$, and $\Delta_2 = -m \Delta_1$. There, we observe a general trend that the magnitudes of both $\nu_0$ and $\nu_\pi$ increase by one whenever $m$ increases by around $\pi$, and the jumps in $\nu_0$ ($\nu_\pi$) occurs in the vicinity of $m=n\pi/2$, where $n$ is an even (odd) integer. In particular, we show explicitly in Appendix~\ref{app1} that in the limit $\mu_1=2J_1=2\Delta_1$, $\nu_0$ and $\nu_\pi$ take integer values, which keep increasing whenever $m$ hits $n\pi/2$. As a result, we can in principle generate as many MZMs and MPMs as we desire by tuning the ratio between the energy scales of the two Hamiltonians $H_1$ and $H_2$. It is also worth noting that the binary driving used to describe our model can instead be simulated with a finite collection of harmonic driving with commensurate frequencies \cite{Shikun}, thus opening up more possibilities for its experimental realization. { Finally, we emphasize that while the many MZMs and MPMs described above relies on chiral symmetry protection, perturbations capable of breaking such a chiral symmetry typically involve terms $\propto \sigma_x$, which correspond to the presence of other types of superconductivities with different phase than $\Delta$ and are expected to be unlikely to occur naturally. Note also that, the use of symmetric time frame of Eq.~(\ref{stf}) is not necessary in the actual implementation of the drive and is merely utilized in the above to construct the two invariants $\nu_0$ and $\nu_\pi$.}

\section{Detection of quasiparticle poisoning}
\label{det}

As outlined in Sec.~\ref{intro}, QP presents a main challenge that hinders the full topological protection of realistic Majorana qubit architectures. It takes place when the platform used to host Majorana fermions is not completely isolated from its surrounding, thus allowing Majorana fermions to flow into and from the system. This is the case for most existing Majorana qubit designs, since topological superconductivity is typically achieved through proximity effect, and as such coupling between such systems with a trivial bulk superconductor is unavoidable.

{ In the following, we adopt a phenomenological error model associated with a QP error, without referring to details about its physical origin.  Rather, we use such a model to focus on the effect of a QP error on our state where quantum information is encoded.} In this case, the presence of a QP error can be represented by the application of a single Majorana operator $\gamma_{i}$ (where $c_i\propto \gamma_{2i}-\mathrm{i} \gamma_{2i+1}$) on the logical qubit state. Most of such Majorana operators do not commute with the system's Hamiltonian and its corresponding Floquet operator, which thus transform the logical qubit state into a new state outside the logical subspace (the subspace generated by the Majorana zero and $\pi$ modes). By ensuring a large gap $\Delta \varepsilon_{0}$ ($\Delta \varepsilon_{\pi/T}$) between quasienergy zero ($\pi/T$) with the rest of the bulk quasienergy excitations, such error processes are expected to be suppressed at low enough temperatures, i.e., when $kT\ll \mathrm{min} \left(\Delta \varepsilon_{0}, \Delta \varepsilon_{\pi/T}\right)$. { Here, ``temperatures" refer to those involved in the preparation of the Floquet state representing our qubit state. In particular, while the notion of ground state is not properly defined in Floquet systems, a Floquet state can in principle be obtained by first starting with a ground state of a certain static system, then adiabatically turning on a periodic-driving to eventually achieve the time-periodic Hamiltonian under consideration \cite{Fprep1,Fprep2}.}

A more dangerous error may occur when such a QP process involves one of the Majorana zero and $\pi$ modes, which causes a logical error within the logical subspace and is thus not protected by a quasienergy gap. Consider for example a periodically driven quantum wire hosting only a pair of MZM and MPM at each of its ends, which can be labelled by $\gamma_{0,L}$, $\gamma_{0,R}$, $\gamma_{\pi,L}$, and $\gamma_{\pi,R}$. The four basis qubit states can then be chosen as $|00\rangle$, $|10\rangle =\gamma_{0,L} |00\rangle$, $|01\rangle =\gamma_{\pi,L} |00\rangle$, and $|11\rangle = \gamma_{0,L} \gamma_{\pi,L} |00\rangle$, where $\mathrm{i} \gamma_{0,L}\gamma_{0,R} |00\rangle = \mathrm{i} \gamma_{\pi,L}\gamma_{\pi,R} |00\rangle = |00\rangle$ \cite{note}, { where the occupations of other (bulk) modes are fixed.} It then follows that the occurrence of even a single QP associated with either $\gamma_{0,L}$, $\gamma_{0,R}$, $\gamma_{\pi,L}$, or $\gamma_{\pi,R}$ transforms one basis qubit state to another, thus generally resulting in an (undetected) erroneous state.

A standard procedure in many quantum error correction (QEC) schemes is to encode a single logical qubit using multiple physical qubits which allows us to utilize the additional degrees of freedom for detecting and subsequently correcting errors \cite{ECR}. In order to implement a QEC protocol capable of resolving QP, it is thus necessary to prepare a system with multiple MFs, which can be accomplished by following our recipe presented in Sec.~\ref{sec2}. In particular, a minimum of $14$ MFs is required to encode a single logical qubit with protection against any QP error. In this case, we may implement the Majorana version of the Steane $[[7,1,3]]$ code \cite{EC,Fu,Ste}.

In the system introduced in Sec.~\ref{sec2}, this is achieved by first tuning the parameter regime at which $\nu_0=3$ and $\nu_\pi=4$ in Fig.~\ref{wmj}, where it hosts six and eight MZMs and MPMs respectively at its ends. We may then label these MZMs and MPMs as $\gamma_{0,S,i}$ and $\gamma_{\pi,S,\alpha}$ respectively, with $S\in \left\lbrace L,R \right\rbrace$, $i\in \left\lbrace 1,2,3 \right\rbrace$, and $\alpha\in \left\lbrace 1,2,3,4 \right\rbrace$, to define the following stabilizer operators,
\begin{eqnarray}
\mathcal{S}_1 &=& \gamma_{0,L,1}\gamma_{0,R,1} \gamma_{\pi,L,1}\gamma_{\pi,R,1}  \;, \nonumber \\
\mathcal{S}_2 &=& \gamma_{0,L,1} \gamma_{0,R,2} \gamma_{\pi,L,1}\gamma_{\pi,R,2}  \;, \nonumber \\
\mathcal{S}_3 &=& \gamma_{0,L,1} \gamma_{0,R,1} \gamma_{\pi,L,2}\gamma_{\pi,R,2}  \;, \nonumber \\
\mathcal{S}_4 &=& \gamma_{0,L,2} \gamma_{0,R,3} \gamma_{\pi,L,3}\gamma_{\pi,R,3}  \;, \nonumber \\
\mathcal{S}_5 &=& \gamma_{0,L,3} \gamma_{0,R,3} \gamma_{\pi,L,3}\gamma_{\pi,R,4}  \;, \nonumber \\
\mathcal{S}_6 &=& \gamma_{0,L,2} \gamma_{0,R,3} \gamma_{\pi,L,4}\gamma_{\pi,R,4}  \;. \nonumber \\
\label{stab}
\end{eqnarray}
It can be easily verified (see also Table.~\ref{tab1}) that each $\gamma_{0,S,i}$ and $\gamma_{\pi,S,\alpha}$ anticommute with a \emph{unique} collection of $S_i$ above while commuting with the rest. As a result, by initializing the logical qubit state in a subspace where all $\mathcal{S}_i=+1$, the occurrence of a single QP process leads to the change in some of the $\mathcal{S}_i$ eigenvalues to $-1$, which can be detected by actively measuring all the stabilizer operators. Correction routine can then be performed by either purposely injecting an appropriate Majorana fermion based on the detected QP event, or simply by utilizing classical computer to reinterpret the stabilizer measurements.

\begin{table}
	\begin{center}
		\item \begin{tabular}{|c|c|}\hline
			\textbf{Error} & \textbf{Anticommutes with} \\\hline
			$\gamma_{0,L,1}$ & $\mathcal{S}_1$, $\mathcal{S}_2$, and $\mathcal{S}_3$ \\ \hline
			$\gamma_{0,L,2}$ & $\mathcal{S}_4$ and $\mathcal{S}_6$ \\ \hline
			$\gamma_{0,L,3}$ & $\mathcal{S}_5$ \\ \hline
			$\gamma_{0,R,1}$ & $\mathcal{S}_1$ and $\mathcal{S}_3$  \\ \hline
			$\gamma_{0,R,2}$ & $\mathcal{S}_2$ \\ \hline
			$\gamma_{0,R,3}$ & $\mathcal{S}_4$, $\mathcal{S}_5$, and $\mathcal{S}_6$ \\ \hline
			$\gamma_{\pi,L,1}$ & $\mathcal{S}_1$, and $\mathcal{S}_2$ \\ \hline
			$\gamma_{\pi,L,2}$ & $\mathcal{S}_3$ \\ \hline
			$\gamma_{\pi,L,3}$ & $\mathcal{S}_4$ and $\mathcal{S}_5$ \\ \hline
			$\gamma_{\pi,L,4}$ & $\mathcal{S}_6$ \\ \hline
			$\gamma_{\pi,R,1}$ & $\mathcal{S}_1$ \\ \hline
			$\gamma_{\pi,R,2}$ & $\mathcal{S}_2$ and $\mathcal{S}_3$ \\ \hline
			$\gamma_{\pi,R,3}$ & $\mathcal{S}_4$ \\ \hline
			$\gamma_{\pi,R,4}$ & $\mathcal{S}_5$ and $\mathcal{S}_6$ \\ \hline
		\end{tabular}
	\end{center}
\caption{A list of all QP errors that anticommute with a unique collection of stabilizer operators.}
\label{tab1}
\end{table}

Since the above error correction has a code distance of $d=3$, it is only capable of successfully detecting and correcting a single QP error as elucidated above. It can be checked however that certain (potentially dangerous) weight-two errors, such as $\mathrm{i} \gamma_{0,L,1}\gamma_{0,L,2}$ or $\mathrm{i} \gamma_{0,L,1}\gamma_{0,L,3}$ which may represent either dephasing errors or the occurrence of two QP events involving two neighboring MFs at the same energy, can also be uniquely corrected. In general, a pair of MFs, one of which anticommutes with some stabilizers in $\left\lbrace \mathcal{S}_1, \mathcal{S}_2, \mathcal{S}_3\right\rbrace$, whereas the other anticommutes with some stabilizers in $\left\lbrace \mathcal{S}_4, \mathcal{S}_5, \mathcal{S}_6\right\rbrace$ are correctable. This can be understood from the fact that these two sets of stabilizers belong to two different copies of $[[7,1,3]]$ codes which are combined together to obtain a physical Majorana fermion code with an even number of Majorana fermions \cite{EC,Fu}, so that such weight-two errors can be interpreted as two weight-one errors occuring separately at each code and are thus correctable.

Finally, it should be noted that the Majorana Steane $[[7,1,3]]$ code elucidated above represents only a minimal code specifically targeted to combat a single QP event. Other more sophisticated error correction codes which are able to fully correct higher-weight errors also exist \cite{Fu}. Such codes necessarily involve more MFs, but this aspect is not an issue for our proposed model given that any number of MFs can be realized by simply tuning some system parameters. Potential challenges in the experimental implementation of these more general codes may nevertheless still arise due to higher-weight MFs involved in the stabilizers and the corresponding logical qubit operators. In this case, a generalization of Majorana Steane code with higher code distance and constant-weight Majorana stabilizers is desirable and may serve as an interesting future work.

\section{Physical implementation}

\subsection{Measurement of stabilizer operators}
\label{condmea}

Having established a minimal stabilizer code based on the multiple MZMs and MPMs at the ends of a single periodically-driven quantum wire, a means to measure the proposed stabilizer operators is an important aspect that we will now address. To this end, we focus our attention on the semiconducting nanowire realization of the periodically quenched Hamiltonian described by Eq.~(\ref{mod}). In particular, a 1D $p$-wave superconductor wire hosting a pair of MZMs at both ends can be realized by proximitizing a semiconducting nanowire having a moderate spin-orbit coupling with a trivial $s$-wave superconducting system, under the application of a perpendicular magnetic field \cite{semi1,semi2}. { While the resulting system may not exactly replicate the periodically driven $p$-wave superconducting wire of Eq.~(\ref{mod}), e.g. given the infinite bandwidth of nanowire systems as compared with the finite bandwidth of Kitaev $p$-wave superconducting model, we expect the two systems to still be qualitatively similar. That is, in addition to the gap opening at quasienergy set by chemical potential due to interplay between $s$-wave superconductivity and magnetic field as typically expected in static systems \cite{semi1,semi2}, additional gaps at quasienergies differing from the chemical potential by $\pm \frac{\hbar \omega}{2}$ are also expected to arise due to resonance between states differing by one photon. In this case, the two gaps allow the MZMs and MPMs expected above to reside.

Explicit dependence of the effective chemical potential and $p$-wave pairing in Eq.~(\ref{mod}) on the magnetic field strength, $s$-wave pairing amplitude, and spin-orbit coupling parameter are discussed in Appendix~\ref{app2}, which suggests that our time-periodic system of Eq.~(\ref{mod}) can be obtained by periodically quenching the hopping amplitude, magnetic field strength, and the chemical potential. In such a nanowire realization of our Floquet topological superconductor, however, the hopping amplitude may not be an easily tunable parameter. Fortunately, additional calculations of the winding number invariants presented in Appendix~\ref{app3} reveal that the presence of many MZMs and MPMs presented earlier (including the case with six MZMs and eight MPMs) can also be achieved by periodically quenching $\mu$ and $\Delta$ (consequently, chemical potential and magnetic field strength) alone. Moreover, although Eq.~(\ref{momH2}) in Appendix~\ref{app2} suggests that the induced $p$-wave pairing can only decrease with the increase in the Zeeman energy and may thus not be able to reach an arbitrarily large value, the required parameter regime at which six MZMs and eight MPMs emerge can still in principle be attained as long as a large ratio between the two tunable parameters, i.e., $\mu_2/\mu_1$ and $\Delta_2/\Delta_1$, be appropriately tuned.}

\begin{figure}
	\centering
	\includegraphics[scale=0.5]{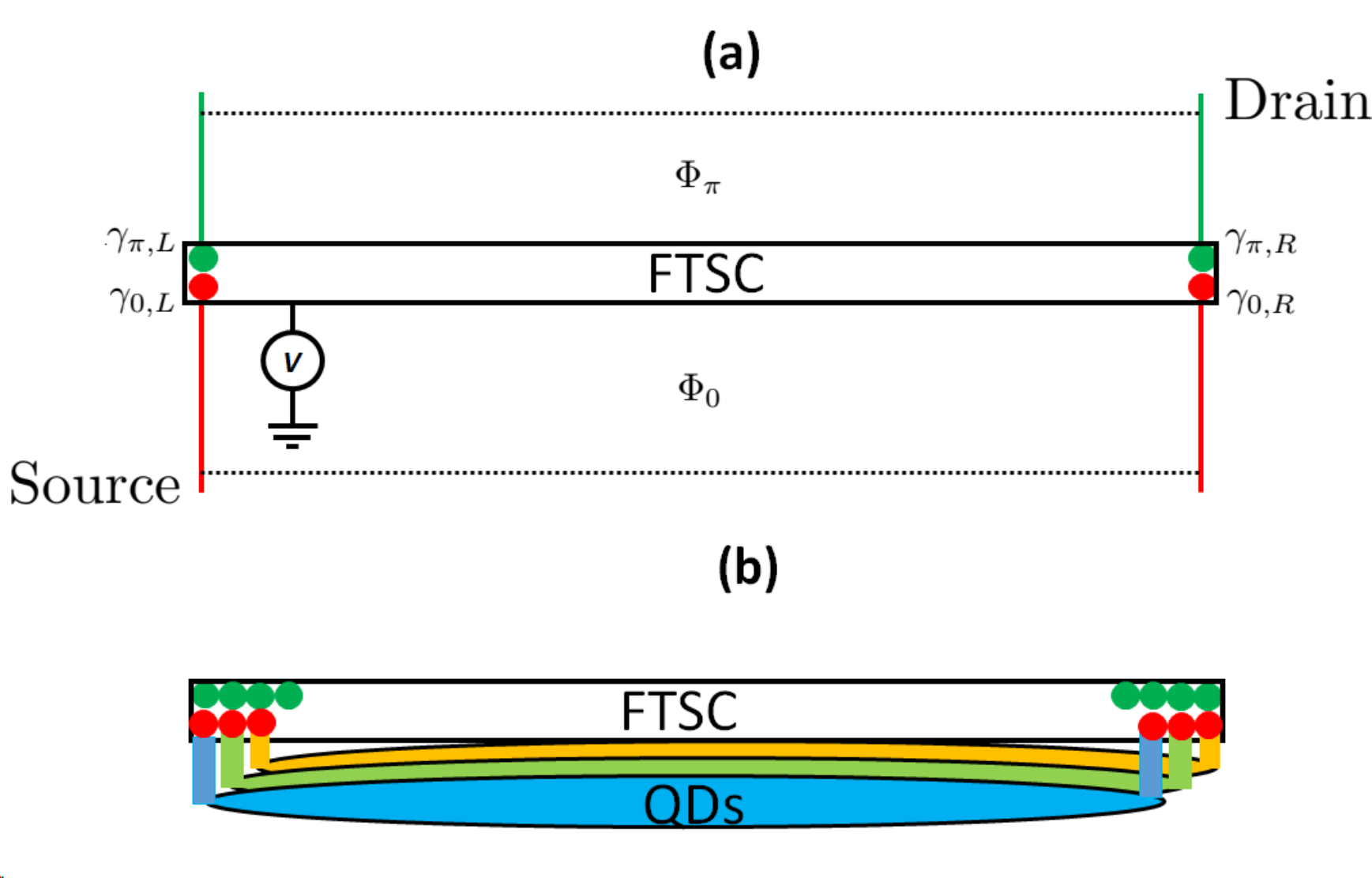}
	\caption{(a) Weight-four Majorana parity can be measured via the parity dependence conductance in a system of Floquet topological superconductor (FTSC) with finite charging energy and four leads attached to its ends. The visibility of such a measurement is controlled by the magnetic flux $\Phi_\pi$ and $\Phi_0$ threading a region between two of the leads. { (b) Measurement of weight-six MZMs via connection to three quantum dots (QDs) that results in parity-dependent quasienergies.}}
	\label{set}
\end{figure}

In the static setup, the parity of its two MZMs manifests itself in the conductance between the two system ends when the latter is connected to a capacitor that maintains a finite charging energy $E_C=\frac{e^2}{2C}$ \cite{mea3}. Our previous work in Ref.~\cite{RG} generalizes such a parity-dependent conductance to capture two- and four-Majorana parities involving MZMs, MPMs, or both. In particular, given a periodically driven topological superconductor hosting in total two MZMs and MPMs, we may attach two external leads at each of its ends. Each of these leads is assumed to be { effectively a single level system (achieved by, e.g., an intermediate connection with a quantum dot)}, where one of them is set at zero energy and the other at $E=\frac{\pi}{T}$. Moreover, a pair of leads at two opoosite ends which have the same energy are weakly connected with a small coupling strength $\lambda_{0(\pi)}$, and a tunable magnetic flux $\Phi_{0(\pi)}$ is threaded into the area spanned by two such leads. The aforementioned setup is summarized in Fig.~\ref{set}(a), where $V$ is the external potential controlling the charging energy $E_C$. Using third order Floquet perturbation theory, whose detail closely follows Ref.~\cite{RG}, the time-averaged conductance between the source and drain leads in Fig.~\ref{set}(a) takes the form
\begin{widetext}
\begin{eqnarray}
\bar{G} &=& a_0 +a_1 \langle \mathrm{i}\gamma_{0,L}\gamma_{0,R}\rangle \sin\left[\frac{e}{\hbar}(\Phi_{0}-\phi_{0})\right] +a_2 \langle \mathrm{i}\gamma_{\pi,L}\gamma_{\pi,R}\rangle \sin\left[\frac{e}{\hbar}(\Phi_{\pi}-\phi_{\pi})\right] \nonumber \\
&& +a_3 \langle \gamma_{0,L}\gamma_{0,R}\gamma_{\pi,L}\gamma_{\pi,R}\rangle \cos\left[\frac{e}{\hbar}(\Phi_{\pi}-\phi_{\pi}-\Phi_{0}+\phi_{0})\right]\;, \label{cond}
\end{eqnarray}	
\end{widetext}
where $a_0$, $a_1$, $a_2$, $a_3$, $\phi_0$, and $\phi_\pi$ depend on the system parameters \cite{RG}. By tuning $\Phi_0\approx \phi_0$ and $\Phi_\pi \approx \phi_\pi$, measuring $\bar{G}$ gives us only one of two possible outcomes which depends on the four-Majorana parity $\langle \gamma_{0,L}\gamma_{0,R}\gamma_{\pi,L}\gamma_{\pi,R}\rangle$. { In particular, fine-tuning $\Phi_0$ and $\Phi_\pi$ to exactly $\phi_0$ and $\phi_\pi$ may not be necessary in practice due to the finite resolution of realistic devices. In this case, as long as the second and third terms of Eq.~(\ref{cond}) are much smaller as compared with its last term, it is in principle possible to measure the desired four-Majorana parity without accidentally measuring the individual two-Majorana parities (e.g. by allowing measurement time to be fast enough to prevent the state from collapsing into an eigenstate of the individual two-Majorana parities, but slow enough to still allow the state to collapse into an eigenstate of the four-Majorana parity).}

\begin{figure}
	\centering
	\includegraphics[scale=0.5]{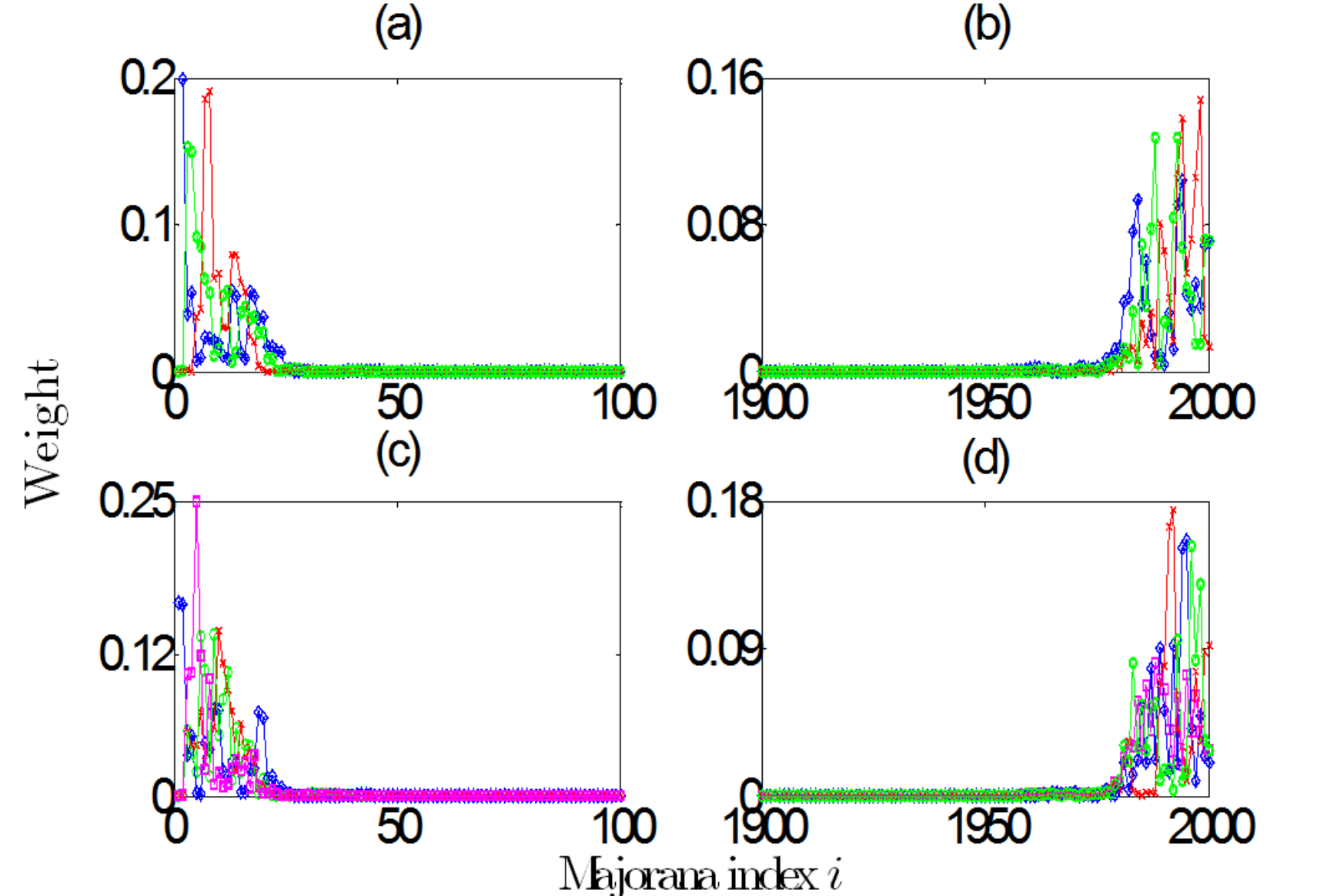}
	\caption{By writing all MZM and MPM solutions in the form of $\gamma= \sum_i w_i \gamma_i$, where $\gamma_i$ represent Majorana operators that are related to the fermion operators as $c_j\propto \gamma_{2j}- \mathrm{i} \gamma_{2j+1}$ and $w_i$ represents the weight (support) of $\gamma$ on Majorana operator $\gamma_i$, the above shows the weight distribution of (a) the three MZMs localized near the left end, (b) the three MZMs localized near the right end, (c) the four MPMs localized near the left end, and (d) the four MPMs localized near the right end. The system parameters are chosen as $\frac{\mu_1 T}{2\hbar}=1$, $\frac{J_1 T}{2\hbar}=\frac{\Delta_1 T}{2\hbar}=0.5$, $\mu_2=m\mu_1$, $J_2=mJ_1$, $\Delta_2=m\Delta_1$, and $m=3.6\pi$.}
	\label{state}
\end{figure}

When there are more than one MZM and MPM at each end of the system, the same setup above can in principle still be implemented to measure certain four-Majorana parity. In particular, since multiple MZMs or MPMs located at the same edge have different supports within the system (see Fig.~\ref{state}), capturing a particular MF, e.g. $\gamma_{0,L,1}$ instead of $\gamma_{0,L,2}$ or $\gamma_{0,L,3}$, in the parity-dependent conductance measurement can be accomplished by appropriately designing the coupling between the lead and the system such that it has a high overlap with certain points in the system which support the desired MF. Given that each stabilizer operator in Eq.~(\ref{stab}) comprises of four MFs with one MZM and one MPM from each end, it can then be measured with the same setup elucidated above. Measurements of different stabilizers then correspond to connecting the four leads at slightly different locations in the system so as to create a large overlap with different MFs that appear in a given stabilizer operator.

\subsection{Logical qubit encoding and scalability}
\label{scale}

The Hilbert space spanned by $14$ MFs and constrained by $6$ stabilizer operators has $2^{\frac{14}{2}-6}=2$ dimensions, which allows the encoding of a single logical qubit. Such a qubit is represented by the Pauli operators $\sigma_z$ and $\sigma_x$ consisting of six and three MFs respectively. For example, we may define $\sigma_z=\mathrm{i} \prod_{S\in \left\lbrace L,R \right\rbrace , i\in \left\lbrace 1,2,3 \right\rbrace} \gamma_{0,S,i}$ and $\sigma_x=\mathrm{i} \gamma_{0,L,1}\gamma_{0,R,1} \gamma_{0,R,2}$, so that $\sigma_z$ and $\sigma_x$ anticommute with each other and commute with all the stabilizer operators $\mathcal{S}_i$. Since $\sigma_z$ comprises of all Majorana zero modes, { it can be measured by connecting the system to three quantum dots (QDs), such that each QD is coupled to a pair of MZMs from both ends (see Fig.~\ref{set}(b)). Following Ref.~\cite{QP2}, this leads to the presence of parity-dependent quasienergies that do not depend on undesired lower-weight Majorana parities, which can then be measured via energy level spectroscopy \cite{QP2}, accomplished by connecting the system to a superconducting transmission line resonator and performing the reflectometry technique to measure the parity-dependent resonator frequency \cite{QP2}.} Together with the first round of all stabilizer measurements, this allows the initialization of a logical qubit state. On the other hand, since $\sigma_x$ involves the product of $3$ MFs, the two logical qubit basis states have different total Majorana fermion parities. A logical $X$-gate can in principle then be implemented by purposely injecting three different species of Majorana fermions onto the system, so as to conserve all the stabilizer operators.

\begin{figure}
	\centering
	\includegraphics[scale=0.5]{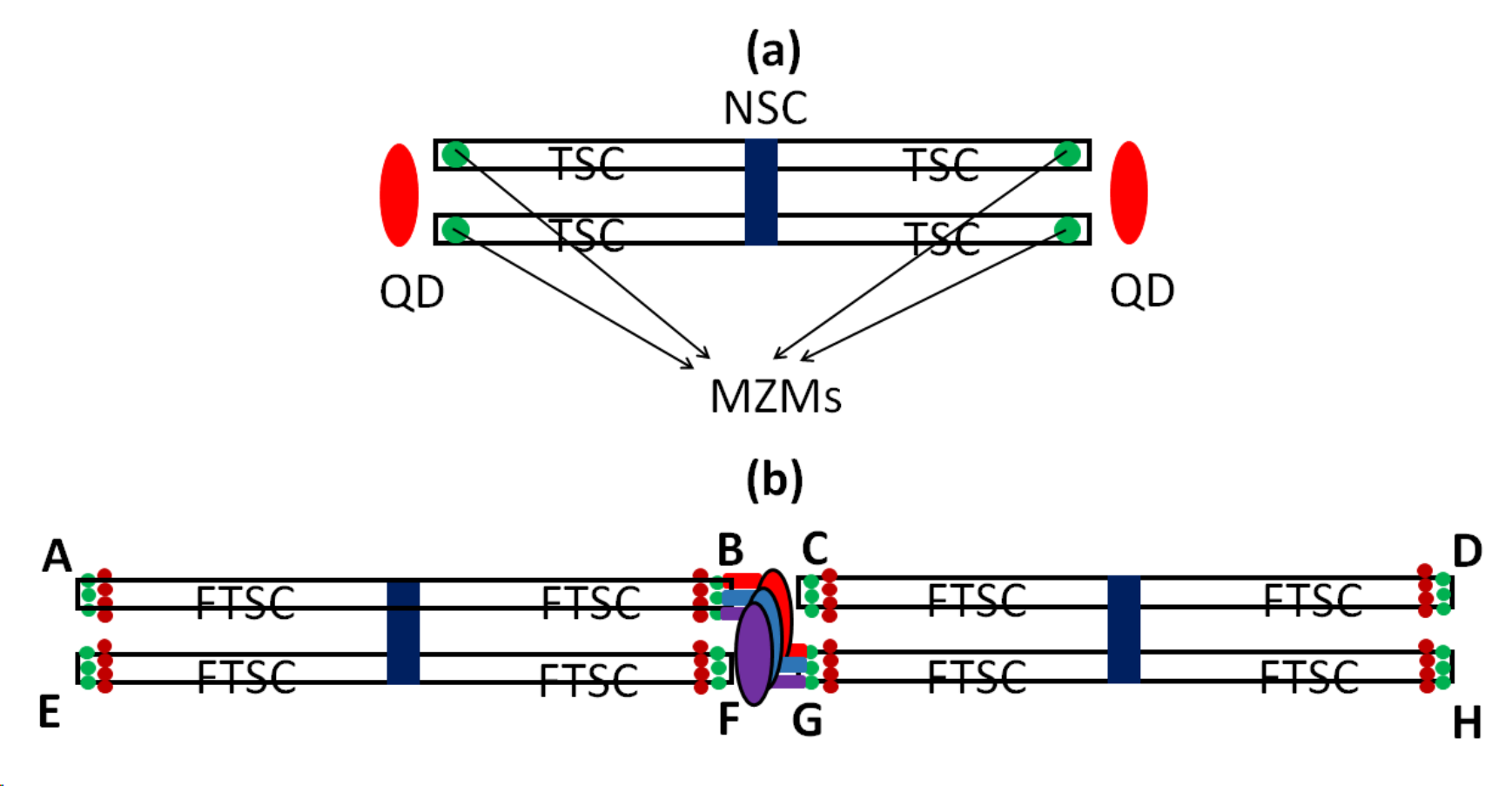}
	\caption{(a) A two-sided tetron architecture introduced in Ref.~\cite{QP2}, which comprises of topological superconductors (TSC), quantum dots (QDs), and a normal superconductor (NSC). Coupling between each MZM and its adjacent QD can be switched on/off via an external gate. (b) A system of two periodically-driven (Floquet) tetrons obtained by periodically driving the TSCs (FTSCs) following the recipe introduced in Sec.~\ref{sec2}. { By turning on the coupling between each MZM at corner B (and G) with the each one of the three QDs}, weight-six Majorana parity measurement of $\mathrm{i}\prod_{S\in\left\lbrace B,G \right\rbrace, i\in\left\lbrace 1,2,3 \right\rbrace} \gamma_{0,i}^S$ (see main text for notation) can in principle be done.}
	\label{ftet}
\end{figure}

While a system of $14$ MFs above serves as a good quantum memory against QP errors and allows the implementation of limited quantum gate operations, it is not sufficient to carry out most quantum computational tasks. In this case, such a system is to be viewed as a building block for incorporating active quantum error correction protocols in various existing qubit architectures, thus providing more protection against QP errors. Indeed, given its capability to host multiple MFs within a single quantum wire, our system is compatible with scalable designs (tetrons and hexons) proposed in Ref.~\cite{QP2}. In particular, all tetrons and hexons variations comprise of arrays of 1D topological superconductors connected together by a normal (backbone) superconductor at one end, so that MFs emerge only at the other end and are adjacent to quantum dots (QDs) \cite{QP2}, whose coupling can be switched on/off via external gates (One of such architectures is illustrated in Fig.~\ref{ftet}(a)).

By driving these 1D topological superconductors according to the recipe introduced in this paper, each of them may now host more than one MFs. Quantum error correction codes such as the Majorana Steane code illustrated above can then be implemented individually within each topological superconductor. In this case, stabilizer measurements via the four-Majorana parity conductance introduced in Sec.~\ref{condmea} can still be carried out through the setup of Fig.~\ref{set}(a), where the associated leads are assumed to be floating. Moreover, since the logical Pauli gates $\sigma_z$ and $\sigma_x$ of these periodically-driven topological superconductors can be chosen to involve only MZMs, it is expected that quantum gate operations with such periodically-driven tetrons and hexons can be performed in almost the exact same way as their static counterparts, i.e., by appropriately turning on some MFs-quantum dot coupling and measuring the parity dependence quantum dot quasienergy.

As an example, Fig.~\ref{ftet}(b) shows a setup comprising of two periodically-driven two-sided tetrons, with four MPMs (brown circles) and three MZMs (green circles) at each end, and { now three quantum dots (colored ellipses)}. By labelling each tetron corner with letters $A$ to $H$, we may label each MZM and MPM as $\gamma_{0,i}^S$ and $\gamma_{\pi,\alpha}^S$, where $i=1,2,3$, $\alpha=1,2,3,4$, and $S=A,B,C,D,E,F,G,H$. Here, each tetron can be assumed to encode a single logical qubit with even-weight MFs Pauli operators, i.e. $\sigma_z^{(1)}=\mathrm{i} \prod_{S\in\left\lbrace B,F \right\rbrace, i\in\left\lbrace 1,2,3 \right\rbrace} \gamma_{0,i}^S$, $\sigma_x^{(1)}=\mathrm{i} \prod_{S\in\left\lbrace E,F \right\rbrace, i\in\left\lbrace 1,2,3 \right\rbrace} \gamma_{0,i}^S$, $\sigma_z^{(2)}=\mathrm{i} \prod_{S\in\left\lbrace C,G \right\rbrace, i\in\left\lbrace 1,2,3 \right\rbrace} \gamma_{0,i}^S$, and $\sigma_x^{(2)}=\mathrm{i} \prod_{S\in\left\lbrace G,H \right\rbrace, i\in\left\lbrace 1,2,3 \right\rbrace} \gamma_{0,i}^S$ for the first and second tetrons respectively. By regarding the second qubit as an ancilla, certain quantum gates can be implemented by a series of weight-six Majorana parity measurements, each of which can be carried out by turning on the coupling between two appropriate corners and the QDs (one example is also illustrated in Fig.~\ref{ftet}). For example, a $\pi/2$-phase gate $P=\frac{1}{\sqrt{2}}(1+\mathrm{i} \sigma_z^{(1)})$ can be realized, up to measurement-outcome dependent correction, by measuring $\mathrm{i} \prod_{S\in\left\lbrace C,G \right\rbrace, i\in\left\lbrace 1,2,3 \right\rbrace} \gamma_{0,i}^S$, $\prod_{S\in\left\lbrace B,G \right\rbrace, i\in\left\lbrace 1,2,3 \right\rbrace} \gamma_{0,i}^S$, $\prod_{S\in\left\lbrace F,G \right\rbrace, i\in\left\lbrace 1,2,3 \right\rbrace} \gamma_{0,i}^S$, and $\prod_{S\in\left\lbrace C,G \right\rbrace, i\in\left\lbrace 1,2,3 \right\rbrace} \gamma_{0,i}^S$ in this exact order. It thus follows that scaling up such periodically-driven tetrons to encode and manipulate multiple qubits straightforwardly follows that of their static counterpart \cite{QP2}, with the advantage that each logical qubit can now be embedded with quantum error correction codes.

\section{Concluding remarks}
\label{conc}

In this paper, we proposed a recipe to realize multiple MFs at the ends of a single topological superconducting wire, which is necessary for the implementation of various quantum error correction codes. A particular example of such codes, i.e., the Majorana Steane $[[7,1,3]]$ code, was highlighted for its capability to correct any singe QP error. Moreover, the stabilizers of such a code involve only weight-four MFs, which can thus be measured via the Floquet generalization of parity-dependent conductance proposed in our previous work \cite{RG}.

Two important features of our system for the implementation of the Majorana Steane $[[7,1,3]]$ code are as follows. First, the whole error correction protocol can in principle be realized with just a single 1D quantum wire. Second, its logical qubit operators involve only MZMs. Consequently, our system can straightforwardly be integrated into scalable qubit architectures proposed in Ref.~\cite{QP2}, essentially equipping them with error correction codes.

To summarize, our work further demonstrates the use of Floquet topological phases for quantum computation, complementing our previous work \cite{braid6,braid7,RG}. In particular, our earlier work \cite{braid6,braid7,RG} focused on utilizing the presence of MPMs to encode and manipulate additional qubits, thus potentially reducing the required physical resources. Here, we take a step forward by showing that a cleverly designed Floquet topological superconductor allows the implementation of quantum error corrections in a minimal setup. Overall, however, these proposals of Floquet quantum computing are still at its earlier stage, and there is room for improvement in future work.

For example, heating is usually considered as one main issue in harnessing the properties of Floquet many-body systems, which causes many of such systems to eventually thermalize to a trivial infinite-temperature-like state. On the other hand, several studies \cite{DTC1,DTC2,DTC3} have also hinted the possibility, at least for some parameter values, that thermalization may be very slow and become significant only at a large time-scale beyond that typically considered in experiment. In this case, a more detailed study for characterizing a thermalization time in the context of Floquet topological superconductors' potential (e.g. semiconductor- or cold atom-based) realization constitutes an important aspect to pursue in the future. In particular, by comparing such a thermalization time with a typical Majorana coherence time that is well-studied in static systems, one may determine if heating is indeed a major issue that needs to be resolved separately, or if it falls within the same problem of extending Majorana coherence time that is still actively studied even in static systems. Even in the former scenario, several proposals exist to combat heating effect, such as via inducing many-body localization \cite{MBL1,MBL2,MBL3,MBL4,MBL5,MBL6,MBL7} or prethermalization states \cite{pretherm}. To this end, designing a Floquet topological superconductor which inherently exhibits these features may also serve as an interesting future work.

Finally, it is also noted that experimental studies of Floquet MFs are presently not as abundant as their static counterpart. Indeed, even an experimental detection of MPMs would constitute a major progress towards the eventual realizations of our various Floquet quantum computation proposals \cite{braid6,braid7,RG}. To this end, it is worth noting that, quantum error correction application aside, our system with multiple MZMs and MPMs above may also be potentially utilized in experiments to more unambiguously identify the signatures of both MZMs and MPMs (thus avoiding potential misidentification of similar quasiparticles such as the Andreev bound states \cite{ABS1,ABS2}).

\begin{acknowledgements}
	{\bf Acknowledgement}: The work by R.W.B is partially supported by the Australian Research Council Centre of Excellence for Engineered Quantum Systems (EQUS, CE170100009). The work by J.G. is funded  by the Singapore NRF
	Grant No. NRF-NRFI2017-04 (WBS No. R-144-000-378-
	281) and by the Singapore Ministry of Education Aca-
	demic Research Fund Tier-3 Grant No. MOE2017-T3-1-
	001 (WBS. No. R-144-000-425-592).
\end{acknowledgements}

\appendix

\section{Analytical calculations of $\nu_0$ and $\nu_\pi$} \label{app1}

To easily understand how the winding numbers $\nu_0$ and $\nu_\pi$ can be systematically increased by tuning the ratio between the energy scales of $H_1$ and $H_2$ (characterized by the parameter $m$), we will restrict our discussion below to the case $\mu_1=J_1=\Delta_1=\delta$ and show the exact locations at which $\nu_0$ or $\nu_\pi$ jumps by exactly one. As we have demonstrated numerically in the main text however, the same qualitative pattern is also observed even at other parameter values, in which case the jumps in $\nu_0$ or $\nu_\pi$ do not occur exactly at the predicted locations below.

With the help of the Euler formula for Pauli matrix $e^{\mathrm{i}\theta \hat{n}\cdot \sigma}=\cos(\theta)+\mathrm{i} \sin(\theta) \hat{n}\cdot \sigma$, we may write $F(k)$ in a canonical basis explicitly as
\begin{widetext}
\begin{eqnarray}
F(k) &\hat{=}&\left( \begin{array}{cc}
\cos(\theta_-)\cos(m\theta_+) -\mathrm{i}\sin(\theta_-)\sin(m\theta_+)  & e^{-\mathrm{i}k/2} \left[\cos(\theta_-)\sin(m\theta_+) +\mathrm{i} \cos(m\theta_+)\sin(\theta_-) \right] \\ -e^{\mathrm{i}k/2} \left[\cos(\theta_-)\sin(m\theta_+) -\mathrm{i} \cos(m\theta_+)\sin(\theta_-) \right] & \cos(\theta_-)\cos(m\theta_+) +\mathrm{i}\sin(\theta_-)\sin(m\theta_+)
\end{array} \right)  \;, \nonumber \\
\theta_\pm &=& \frac{\delta T}{4\hbar} \sqrt{2(1\pm \cos(k))}
\end{eqnarray}
\end{widetext}
Let us now define $z=\theta_- + \mathrm{i} m \theta_+$. This allows us to write the winding numbers as contour integrations
\begin{widetext}	
\begin{eqnarray}
\nu_0 &=& -\frac{1}{2} -\frac{1}{4\pi\mathrm{i}} \oint \frac{\sin(\mathrm{Re}(z))\sin(\mathrm{Im}(z)) dz+ \mathrm{i} \cos(\mathrm{Re}(z))\cos(\mathrm{Im}(z)) dz^*}{\cos(\mathrm{Re}(z))\sin(\mathrm{Im}(z)) +\mathrm{i} \cos(\mathrm{Im}(z))\sin(\mathrm{Re}(z))} \;, \nonumber \\
\nu_\pi &=& -\frac{1}{4\pi \mathrm{i}} \oint \frac{\sin(\mathrm{Re}(z))\cos(\mathrm{Im}(z))dz^*-\mathrm{i} \cos(\mathrm{Re}(z))\sin(\mathrm{Im}(z)) dz}{\cos(\mathrm{Re}(z))\cos(\mathrm{Im}(z)) +\mathrm{i} \sin(\mathrm{Im}(z))\sin(\mathrm{Re}(z))} \;,
\end{eqnarray}
\end{widetext}
where the first term in the right hand side of $\nu_0$ comes from the winding number of $e^{-\mathrm{i} k/2}$ and the additional $1/2$ factor in the contour integration comes from the fact that $z$ only traverses half a loop when $k$ is varied from $-\pi$ to $\pi$.

The contour integrations in $\nu_0$ and $\nu_\pi$ can be carried out using Cauchy residue theorem. To this end, let $n\pi/2<m<n\pi/2+\pi$, where $n$ is an odd integer. It then follows that the contour integration, as parametrized by $k$, encloses the points $z_{\pm l}=0\pm\mathrm{i} l \pi/2$, where $l$ is an odd integer $\leq n$. These points correspond to the poles of the integrand inside $\nu_\pi$, where their residue can be calculated via the formula
\begin{equation}
\mathrm{Res}_{z_{\pm l}}\left(\nu_\pi \right) =\lim\limits_{z\rightarrow z_{\pm l}} \mathrm{i} (z-z_{\pm l}) \frac{\cos(x)\sin(y)}{f(x,y)}\;,
\end{equation}
where $x=\mathrm{Re}(z)$, $y=\mathrm{Im}(z)$, and $f(x,y)$ is the denominator of the integrand inside $\nu_\pi$. Since $f(x,y)$ is complex differentiable at $z_{\pm l}$, which can be checked via the Cauchy-Riemann equation,
\begin{eqnarray}
\frac{\partial \mathrm{Re}(f)}{\partial x} &=& \frac{\partial \mathrm{Im}(f)}{\partial y} = 0\;, \nonumber \\
\frac{\partial \mathrm{Re}(f)}{\partial y} &=& -\frac{\partial \mathrm{Im}(f)}{\partial x} = - \cos(\mathrm{Re}(z))\sin(\mathrm{Im}(z)) \;, \nonumber \\
\end{eqnarray}
we can evaluate the limit using L'Hopital's rule to arrive at
$\mathrm{Res}_{z_{\pm l}}\left(\nu_\pi \right) = 1$. Since there are in total $n+1$ such poles, Cauchy residue theorem gives
\begin{equation}
\nu_\pi = \frac{1}{2}\sum_{s\in\left\lbrace -n,-n+2\cdots , n\right\rbrace} \mathrm{Res}_{z_s}(\nu_\pi) = \frac{n+1}{2} \;.
\end{equation}

The winding number $\nu_0$ can be calculated in a similar fashion. To this end, we now assume that $n\pi< m < n\pi +\pi$, where $n$ is \emph{any} integer. There are in total $2n+1$ poles located at $z_{\pm l}= 0 \pm \mathrm{i} l\pi$ for any nonnegative integer $0\leq l \leq n$. Upon changing the integration variable $z^*\rightarrow z$, we can evaluate the residue as
\begin{equation}
\mathrm{Res}_{z_{\pm l}}\left(\nu_0 \right) =\lim\limits_{z\rightarrow z_{\pm l}} \mathrm{i} (z-z_{\pm l}) \frac{\cos(x)\cos(y)}{g(x,y)}\;,
\end{equation}
where $g(x,y)=-\cos(x)\sin(y)+\mathrm{i} \cos(y)\sin(x)$, which is complex differentiable at $z_{\pm l}$. Applying L'Hopital's rule again leads to $\mathrm{Res}_{z_{\pm l}}\left(\nu_0 \right)=1$, thus implying
\begin{equation}
\nu_0 = -\frac{1}{2} +\frac{1}{2} \sum_{s\in\left\lbrace -n,-n+1\cdots , n\right\rbrace} \mathrm{Res}_{z_s}(\nu_\pi) = n \;.
\end{equation}

\section{Effective $p$-wave superconductivity in a proximitized semiconductor system}
\label{app2}

Here, we summarize the main idea of Refs.~\cite{semi1,semi2} (see also \cite{sum}) to elucidate how $p$-wave superconductivity can be realized in a semiconducting nanowire proximitized by a trivial $s$-wave superconductor. Without loss of generality, we may assume that the nanowire is aligned in the $y$-direction, and is subject to an external magnetic field in the $x$-direction. Its total Hamiltonian can then be written as
\begin{eqnarray}
H &=& \sum_{j,\sigma} \left[\frac{\mu}{2} d_{j,\sigma}^\dagger d_{j,\sigma} -J d_{j+1,\sigma}^\dagger d_{j,\sigma} +  \mathrm{i}  \sigma\alpha d_{j+1,\sigma}^\dagger d_{j,\sigma}\right. \nonumber \\
&& \left. + \frac{E_Z}{2} d_{j,\sigma}^\dagger d_{j,\bar{\sigma}} +\sigma \frac{\Delta_s}{2} d_{j,\sigma}^\dagger d_{j,\bar{\sigma}}^\dagger +h.c. \right] \;,
\end{eqnarray}
where $\mu$ is the chemical potential, $J$ is the hopping amplitude, $\alpha$ is the spin orbit coupling strength, $E_Z$ is the Zeeman energy associated with the external magnetic field, $\Delta_s$ is the induced $s$-wave superconductivity, $\sigma=\pm 1$, and $\bar{\sigma}=-\sigma$.

Under PBC, we may define the momentum space operators $d_{k,\sigma} = \sum_j d_{j,\sigma} e^{-\mathrm{i} k j}$ to write the above Hamiltonian as
\begin{eqnarray}
H &=& \sum_{k,\sigma} \left[ \mu d_{k,\sigma}^\dagger d_{k,\sigma} -2 J \cos(k) d_{k,\sigma}^\dagger d_{k,\sigma} +2 \sigma \alpha \sin(k) d_{k,\sigma}^\dagger d_{k,\sigma} \right. \nonumber \\
&& + \left. E_Z d_{k,\sigma}^\dagger d_{k,\bar{\sigma}} + \left( \sigma \frac{\Delta_s}{2} d_{k,\sigma}^\dagger d_{-k,\bar{\sigma}}^\dagger +h.c. \right) \right] \;. \label{momH}
\end{eqnarray}
We may now introduce helical fermion operators \cite{sum}
\begin{eqnarray}
c_{k,\pm} &=& C_\pm d_{k,+} \pm C_\mp d_{k,-} \;, \nonumber \\
C_\pm &=& \sqrt{\frac{1}{2} \pm \frac{\alpha \sin(k)}{\sqrt{(2\alpha \sin(k))^2+E_Z^2}}}\;,
\end{eqnarray}
obtained by diagonalizing the Zeeman and spin-orbit coupling terms of the Hamiltonian. It can easily be checked that $c_{k,\pm}$ indeed satisfies the fermionic commutation relations $\left\lbrace c_{k,s}, c_{k',s'} \right\rbrace =0 $ and $\left\lbrace c_{k,s}^\dagger, c_{k',s'} \right\rbrace =\delta_{k,k'} \delta_{s,s'} $. Moreover, it transforms Eq.~(\ref{momH}) to
\begin{eqnarray}
H &=& \sum_{k,\sigma} \left[ \left(\mu -2J\cos(k) + \sigma \sqrt{(2\alpha \sin(k))^2+E_Z^2} \right) c_{k,\sigma}^\dagger c_{k,\sigma}\right. \nonumber \\
&& + \left. \left( \Delta^{(P)} \sin(k) c_{k,\sigma}^\dagger c_{-k,\sigma}^\dagger + \Delta^{(S)} c_{k,\sigma}^\dagger c_{-k,\bar{\sigma}}^\dagger + h.c. \right)  \right] \;,
\label{effham} \nonumber \\
\end{eqnarray}
where the effective $s$- and $p$-wave pairing in this basis are
\begin{eqnarray}
\Delta^{(S)} &=& \frac{\Delta_s E_Z}{\sqrt{(2\alpha \sin(k))^2+E_Z^2}} \;, \nonumber \\
\Delta^{(P)} &=& \frac{\Delta_s \alpha }{\sqrt{(2\alpha \sin(k))^2+E_Z^2}} \;. \label{momH2}
\end{eqnarray}
It then follows that Hamiltonian of the form Eq.~(\ref{mod}) is obtained by projecting out one of the helical fermions $c_{k,\pm}$ (e.g. by setting $c_{k,+}=c_{k,+}^\dagger = 0$ in Eq.~(\ref{effham})), which is valid in the limit $E_Z\gg \alpha, \Delta_s$ where the $\pm$ helical branches are well separated in energy.

{
\section{Additional numerical results of winding numbers}
\label{app3}

The analytical calculations of the winding numbers presented in Appendix~\ref{app1}, which systematically demonstrate how arbitrarily many MZMs and MPMs can be obtained, are based on periodically-modulating the three parameters $\mu$, $J$, and $\Delta$. Here, we present additional numerical calculations of $\nu_0$ and $\nu_\pi$ when only $\mu$ and $\Delta$ are periodically modulated. In Fig.~\ref{wmj2}, we plot the invariants $\nu_0$ and $\nu_\pi$ by setting $\mu_2=-m\mu_1$, $\Delta_2=-m\Delta_1$, and $J_1=J_2$. It follows that while $\nu_0$ and $\nu_\pi$ are no longer monotonically increasing with the energy scale ratio $m$, they may still eventually become bigger as $m$ keeps increasing. In particular the regime with $6$ MZMs and $8$ MPMs can also be obtained by setting $m/\pi$ between $6$ and $7$.

\begin{figure}
	\centering
	\includegraphics[scale=0.25]{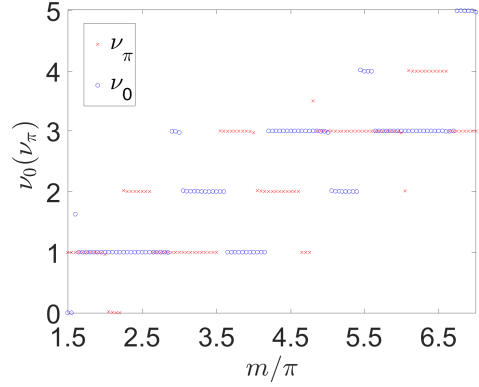}
	\caption{Winding number invariants $\nu_0$ and $\nu_\pi$ obtained by periodically modulating system parameters $\mu$ and $\Delta$.}
	\label{wmj2}
\end{figure}

}


\begin{thebibliography}{99}
\bibitem{note} In the absence of QP errors, the total fermion parity $\gamma_{0,L}\gamma_{0,R}\gamma_{\pi,L}\gamma_{\pi,R}$ is conserved. As a result, each basis qubit state belongs to one of two distinct superselection sectors $\left\lbrace |00\rangle , |11\rangle\right\rbrace$ and $\left\lbrace |01\rangle , |10\rangle\right\rbrace$, and the system effectively encodes only one logical qubit.
\end{thebibliography}
\end{document}